# Increasing the Fine Structure Visibility of the Hinode SOT Ca II H Filtergrams.

E. Tavabi[a, b] and S. Koutchmy[b]

[a] physics Department, Payame Noor University, 19395-4697 Tehran, I, R. of Iran,

[b] Institut d'Astrophysique de Paris, UMR 7095 CNRS & UPMC, 98 Bis Boulevard Arago, F-75014 Paris, France

(E-mail:, tavabi@iap.fr and koutchmy@iap.fr).

**Abstract**

We present the improved so-called Madmax (OMC) operator selecting maxima of convexities computed in multiple directions around each pixel rewritten in MatLab and shown to be very efficient for pattern recognition.

The aim of the algorithm is to trace the bright hair-like features (for ex. chromospheric thin jets or spicules) of solar ultimate observations polluted by a noise of different origins. This popular spatial operator uses the second derivative in the optimally selected direction for which its absolute value has a maximum value. Accordingly, it uses the positivity of the resulting intensity signal affected by a superposed noise. The results are illustrated using a test artificially generated image and real SOT (Hinode) images are also used, to make your own choice of the sensitive parameters to use in improving the visibility of images.

1. Introduction

Solar chromospheric modern observations in Ca II H line (and also in a wide range of other rather cool optical lines such as *Hα, Hβ, H & K of Ca II* and also in EUV lines) exhibit a myriad of thin and highly dynamical features usually called limb spicules. They can reach heights of 4000-10,000km before fading out of view or falling back towards the solar surface. Their smallest widths are only 100-200 *km,* see Tavabi et al. 2011 who statistically analyzed SOT (Hinode) spicules starting from the 1 Mm heights above the limb and found that indeed spicules show a whole range of diameters and time variations, including "interacting spicules" (I-S), depending of the definition chosen to characterize this ubiquitous dynamical phenomenon occurring into a low coronal and chromospheric surroundings.

Almost all spicules seen at the largest heights fade out and/or show a lower contrast above the background than the near limb spicules, although we freely use the term diameter in discussing spicule thickness just because it is convenient to assume that the cross section is circular like in the approximation of a magnetic flux tube.

We first correct the data for dark current and camera artifacts using the IDL routine fg prep, which is a part of the Hinode tree of solarsoft "Level-0" images from the *Hinode* database. This produces Level-1 images corrected for pixel shifts, dark pedestal and current, and flat-field non-uniformity for each filtergram in the data set. Following the fg_prep processing, high-frequency spatial structure is enhanced for the Ca II H-line images by deconvoluting with an experimentally determined instrument point-spread function (PSF) kernel (Liu et al. 2011).

Of course, the background subtraction procedure is not perfect, and residual non-spicule emission may be present in the background-subtracted image or vis. versa.

2. Proposed Algorithms to evaluate images

Recall the 2D Wiener filter, which is optimal in terms of the mean square error, or in other words, which minimizes the overall mean square error in the process of filtering and noise smoothing. The algorithm we discuss is based on the application of an operator pixelwise adaptive method which computes the maxima of convexities in several directions around each pixel and it is shown to be efficient also for pattern recognition. Recall also the popular operator which was advocated in the Rosenfeld and Kak 1976 reference book; it is the 2D Laplacian.

However, the most frequently used method for decreasing a high brightness gradient is the unsharp masking. This method can be understood as a convolution with the derivative of a 2D Gaussian function which enhances the visibility of thin structures, equivalent to a 2D Fourier filtering and it is very useful for evaluating W-L coronal eclipse images see November and Koutchmy, 1986. There exist a large number of numerical, automated codes, specialized either for the case of curvilinear features (Ajabshirizadeh et al. 2008, Aschwanden et al.2008 and Biskri et al. 2010), but there is not a general pattern tracing code which has a superior performance in all kinds of data and features, so for a particular data set and structure it needs to be customized to the particular morphological properties of a given data set, noting also that the manual tracing are not suitable when a large population of similar structures are analyzed.

Accordingly, another operator we prefer, of a more general purpose, is the multi-directional (8, 16, …) maximum gradient operator OMC (Operator Maximum Convexities) used to increase the visibility and contrast of solar structures introduced by Koutchmy and Koutchmy 1989 and also called and nicknamed "Madmax". It has been repeatedly shown that this operator works successfully for improving the visibility of coronal filtergrams, see an excellent example in Takeda et al. 2000, Suematsu et al. 2008 and more recently Anan et al. 2010.

This algorithm is briefly summarized below, for 8 directions and it could be expanded to more radial directions (November and Koutchmy, 1996, Christopoulou et al. 2001).

Let *u(x,y)* be the matrix of measured intensity, the operator M is defined by:

$$M(u(x,y)) = \alpha M_3(u(x,y)) + \beta E_2\left(M_8(S_2(u(x,y)))\right) + \gamma E_4(M_8(S_2(S_2(u(x,y))))) + \ldots \quad (1\text{-}1)$$

Where

$$M_8 = max(-\nabla^2 k) \quad (1\text{-}2)$$

With $\nabla^2 k$ the second order difference operator in k-direction defined by the 0, 45, 90,.. and $S_2$ is the operator which averages every 2 rows and column. $E_2$ and $E_4$ are the operators which expand a matrix by bilinear interpolation with the row and column expansion factor equal to 2 and 4 respectively, α, β and γ are some weights taking arbitrary values.

The most important operator is then $M_8$. Its effect is to amplify convex points, so its main purpose is to find the direction of discontinuity, i.e., locations where there is an abrupt change in gray level. Note that $M_8$ is shift-invariant and is non-idempotent, accordingly it is NOT a morphological filter due to its non linear property. It is less sensitive to an additive noise, see Koutchmy and Koutchmy, 1989.

### 3. Test Result, Applications and Discussion

The OMC spatial filter has been applied to a test 2D texture lines picture (Fig. 1-A) blurred by a Gaussian mask matching an isotropic PSF (point spread function). The FWHM of the Gaussian blurring function is selected to be about 2.5 times the width of the average value of the finest lines thickness see Fig. 1(B).

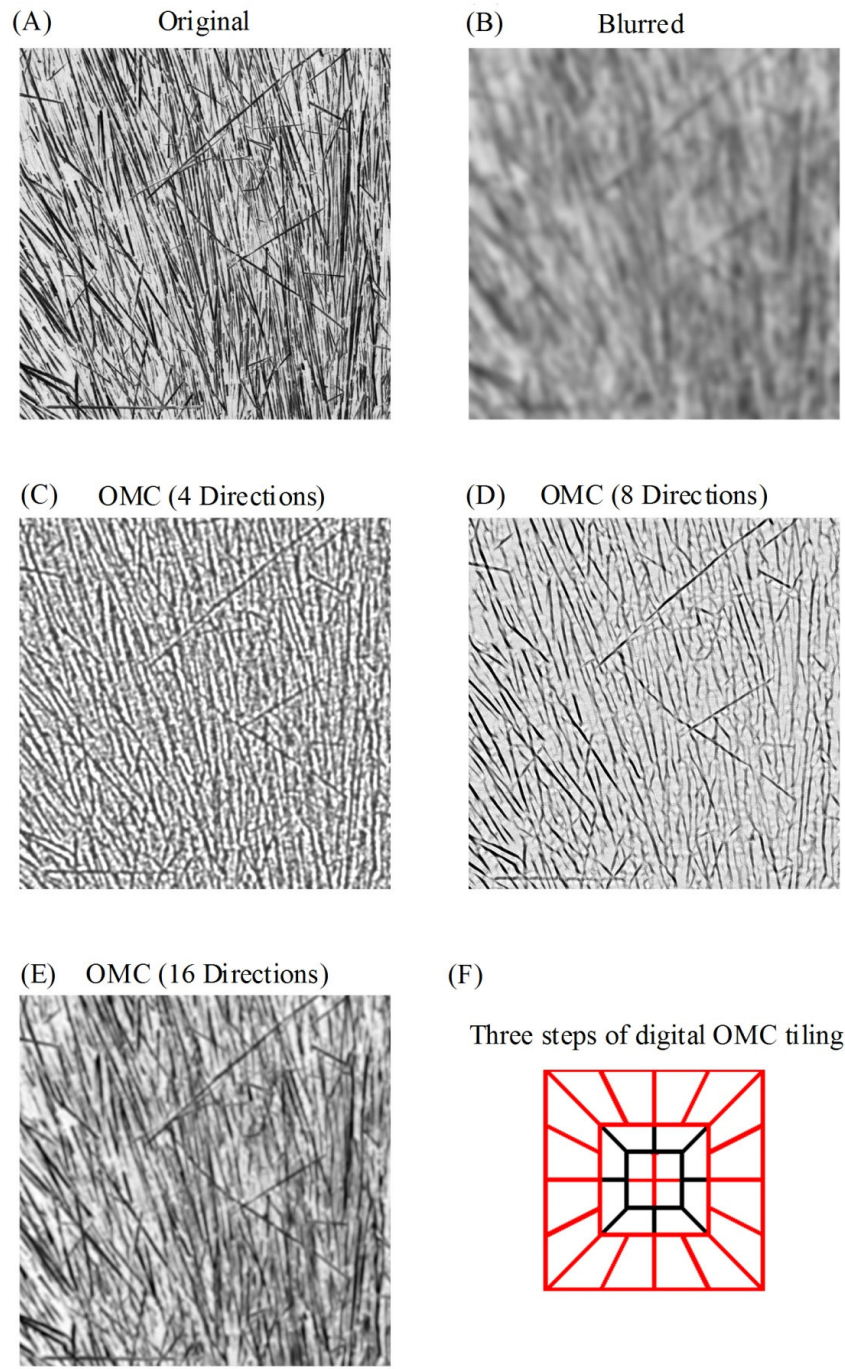

**Fig. 1** Original test image with textures, the corresponding blurred image and the reconstructed using the OMC code (Madmax) applied to the blurred image, for three different steps of tiling.

The result of applying the spatial operator on the visibility and the tracing of hair-like features from a test blurred artificially generated picture (fig. 1-b) is shown in fig. 1-c. This image has been reconstructed first using the second derivative spatial filter in four directions (up, down, left and right, see fig. 1 right bottom panel center section). The four directions second derivative maximum absolute value method is very suitable for edge detection but the finest features are almost removed and a lot of them are definitely lost.

Panels d and f show results for 8 and 16 directions respectively. In eight directions the operator performs well and almost all threads (parallel and crossing) is clearly seen, but at the very location of crossing features, an artificial effect causes a discontinuity along the threads.

In fig. 1-f an even stronger selection is illustrated where the number of directions has been increased to 16; the artificial discontinuities are almost removed and the OMC (Madmax) code looks more optimal: the obtained image has a high correlation with the original un-blurred image. However, more computer resources are needed and this should be considered when processing a sequence of images.

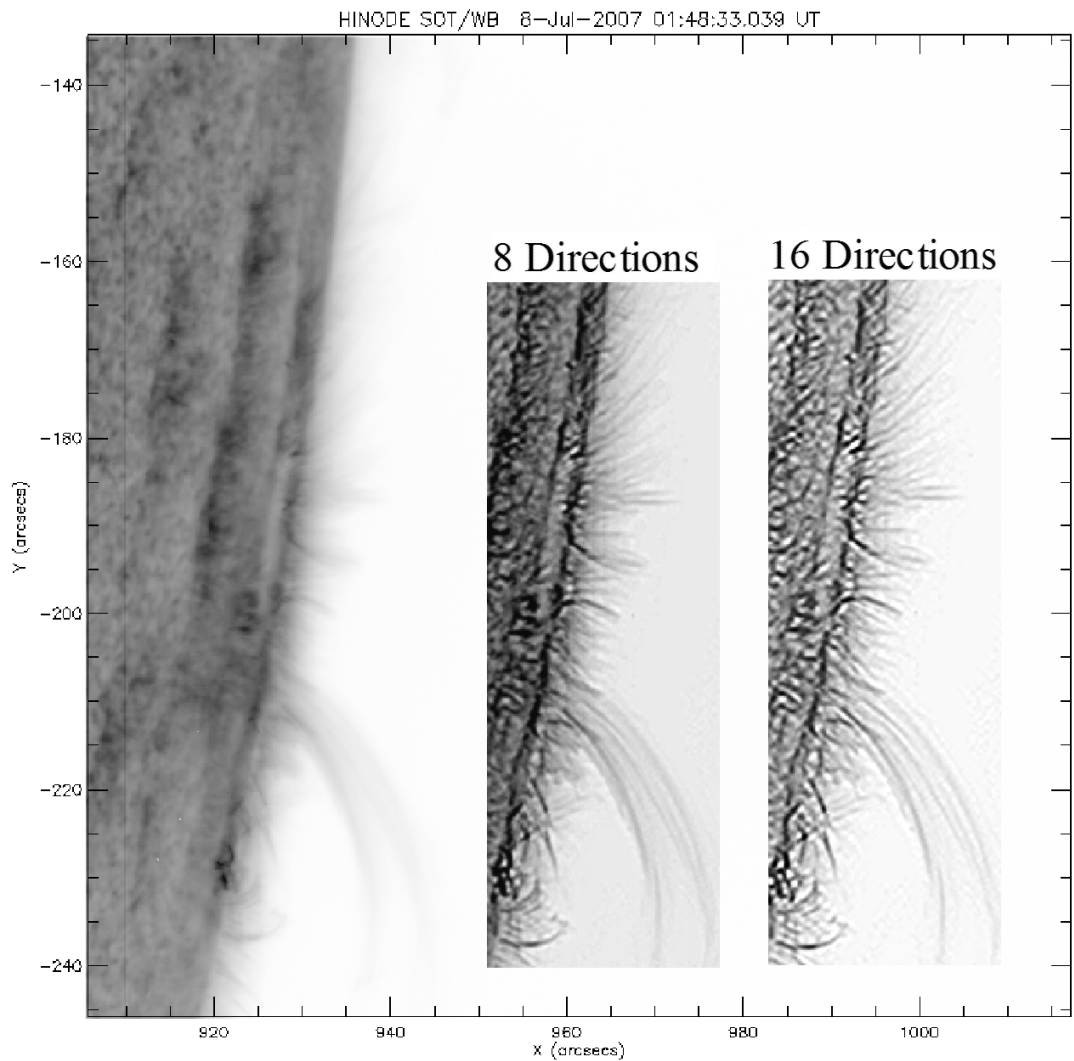

**Fig. 2** From left to right, inverted and calibrated image using the Hinode fg_preb SSW program, and the OMC (Madmax) reconstructed images in 8 and in 16 directions.

Figure 2 shows the use of OMC (Madmax) on a typical beautiful image from a sequence taken with the SOT of Hinode (Shimizu et al. 2008 and Tsunata 2008) using the Ca II H line: a large scale cool jet is ejected from several underlying small loops; they consist of many long parallel threads, all these fine threads being inside the jet only appeared after using the OMC operation, the right frames being the result of applying the algorithm in 8 and 16 directions respectively. The original picture were preprocessed and calibrated using the standard 'fg_prep' of the SSW software.

The OMC (Madmax) tracing method permitted us to look at the feet of the hair-like structures (spicules) above the solar limb. However, looking at the spicule feet suffered from the overlapping effect. The overlapping problem can be understood by calculating the length L along a line of sight to the limb from the tangential point to the height h.

From Pythagoras we have:

$$L^2 + R_0^2 = R_0^2 + 2hR_0 + h^2, \qquad (3\text{-}1)$$

if we neglect $h^2$ as $h \ll R_0$, the double path (in front and back of the limb) is

$$L = \sqrt{(2 R_0 h)}, \qquad (3\text{-}2)$$

All very short chromospheric structures of h=1 Mm are projected against the limb if they are up to 37 Mm along the line of sight, in front or behind the plane of the limb (Zirin 1988).

In case spicules have an average height of 10 Mm, they would cover a 2x186 Mm path which is more than 1 solar radius, so these finely distributed jets will appear as a thick forest!

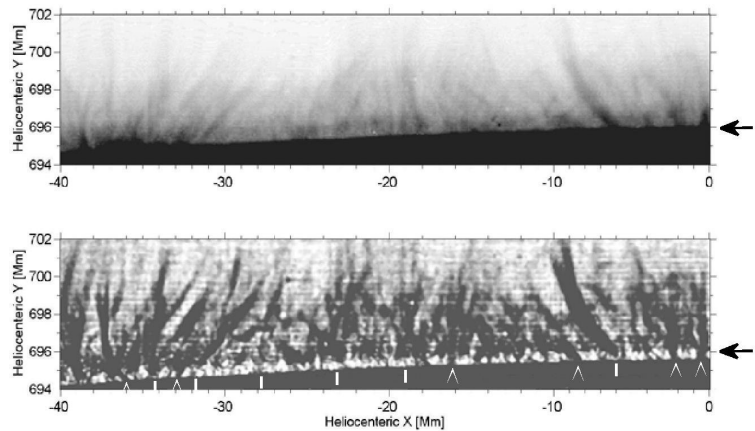

**Fig.3** HCaII SOT- Hinode limb original reconstructed negative, and the corresponding OMC (Madmax) filtergram obtained by superposing and aligning individual frames taken during 5 min at an 8 sec cadence. Arrows point to the same location right above the limb where the image processing performs very efficiently.

As illustrated in Fig. 3, the OMC (Madmax) tracing method permits us to look, for the first time, at the feet of the hair-like structures right above the solar limb where the surrounding or the background atmosphere attenuates the contrast due to the absorption of the line emission of the structure along the line of sight, especially for a spicule situated far behind the limb. It also offers a method to reduce the overlapping effect.

In fig.3, note at the right of each image the arrows placed exactly at the same height to point out the "work" done with the OMC algorithm in showing very near limb structures. At the bottom of the OMC image, under the limb, some marks are put to help to recognize structures supposed to come from behind "Λ" or from beyond "|" the relatively opaque photospheric limb.

On the original image (at the top) the roots of chromospheric structures completely disappear under the 1 Mm heights above the limb, due to a decreasing contrast produced by the effect of i/ the overlapping of several structures along the line of sight and ii/ the increasing diffuse background recorded outside the photospheric limb with both emission effect and absorption effect where a structure is crossed along the line of sight. Note that indeed the disk is very bright because the CaII H filter is broad compared to the width of the CaII H emission line of spicules. Accordingly, in addition to the true background chromospheric diffuse emission outside the limb, some spurious parasitic light of instrumental origin is present. Using Madmaxed images it has been possible to look much deeper towards the limb, under the 1 Mm heights, and see the roots of spicules before and after the limb plane. At the very bottom of what is detected (look at the marks I and ^), approximately half of roots of spicules situated behind the limb is "progressively" disappearing before the 2d half, which is seen significantly lower (closer to the photospheric limb). A special discussion of what is seen, including the small loops very close to the limb and the radiative transfer evaluation, is beyond the scope of this paper. A last point is worth noting: the illustrated processing convincingly shows that it is now possible to improve the hydrostatic homogeneous models of the low chromospheric layers by introducing effects due to the root of spicules with properties definitely different from the homogeneous background. This is also illustrated by the movie made using Madmaxed images showing new dynamical properties of this layer, see Tavabi et al. 2011.

**Acknowledgements.** We are grateful to the Hinode team for their wonderful observations. Hinode is a Japanese mission developed and launched by ISAS/JAXA, with NAOJ as domestic partner and NASA, ESA and STFC (UK) as international partners. This work has been supported by Center for International Scientific Studies & Collaboration (CISSC) and French Embassy in Tehran. The image processing OMC software (often called Madmax) written for IDL is easily downloadable from the O. Koutchmy site at UPMC, see http://www.ann.jussieu.fr/~koutchmy/debruitage/madmax.pro

References

Ajabshirizadeh, A., Tavabi, E., Koutchmy, S.. 2008, New Astronomy **13**, 93-97.

Anan, Tetsu; Kitai, Reizaburo; Kawate, Tomoko; Matsumoto, Takuma; Ichimoto, Kiyoshi; Shibata, Kazunari; Hillier, Andrew; Otsuji, Kenichi; Watanabe, Hiroko; Ueno, Satoru; and 7 coauthors, 2010, PASJ, **62**, 877.

Aschwanden, M.J., Lee, J.K., Gary, G.A., Smith,M., Inhester, B.: 2008, *Solar Phys.* **248**, 359 . doi:10.1007/s11207-007-9064-9.

Biskri, S., Antoine, J.-P., Inhester, B., Mekideche, F., 2010, Sol. Phys., **262**, 373.

Christopoulou, E. B., Georgakilas, A. A., Koutchmy, S., 2001, Sol. Phys., **199**, 61.

Koutchmy, O. & Koutchmy, S., 1989, in Proc. 10$^{th}$ Sacramento Peak Summer Workshop, in High Spatial Resolution Solar Observations, ed. O. von der Luhe (Sunspot: NSO), 217

Liu, Wei, Berger, Thomas E., Title, Alan M., Tarbell, Theodore D., Low, B. C., 2011, ApJ, **728**, 103.

November, L. and Koutchmy, S. 1996, ApJ, **466**, 512

Shimizu, T., Nagata, S., Tsuneta, S., et al., 2008, Sol. Phys., **249**, 22

Suematsu, Y., Ichimoto, K., Katsukawa, Y., Shimizu, T., Okamoto, T., Tsuneta, S., Tarbell, T., Shine, R. A., Astronomical Society of the Pacific, 2008, **397**, 27

Rosenfeld, A. and Kak, A.C. 1976, Digital Picture Processing Academic Press, New York, p. 275.

Takeda, A. Kurokawa, H. Kitai, R. and Ishiura, 2000, PASJ, **52**, 375.

Tavabi, E., Koutchmy, S. and Ajabshirzadeh, A., 2011, New Astronomy, **16**, 296.

Tsuneta, S., Ichimoto, K., Katsukawa, Y., et al., 2008, Sol. Phys., **249**, 167.

Zirin, H., Astrophysics of the Sun, 1988, Cambridge University Press.